\documentclass[11pt,twoside]{article}
\usepackage{asp2014}

\begin{document}

\title{Infrared Opacities in Dense Atmospheres of Cool White Dwarf Stars}

\author{
Piotr M. Kowalski$^1$, Simon~Blouin$^{2,1}$ and Patrick~Dufour$^2$} 
\affil{$^1$IEK-6 Institute of Energy and Climate Research,
Forschungszentrum J\"ulich, 52425 J\"ulich, Germany;}
\affil{$^2$Universit\'e de Montr\'eal, Montr\'eal, Qu\'ebec, Canada}

\begin{abstract}

Dense, He-rich atmospheres of cool white dwarfs represent a challenge to the
modeling. This is because these atmospheres are constituted of a dense fluid in which strong multi-atomic
interactions determine their physics and chemistry. Therefore,
the ideal-gas-based description of absorption is no longer adequate, which makes
the opacities of these atmospheres difficult to model. 
This is illustrated with severe problems in fitting the spectra of cool, He-rich stars.
Good description of the infrared (IR) opacity is essential for proper assignment of the
atmospheric parameters of these stars. Using methods of computational quantum chemistry we
simulate the IR absorption of dense He/H media. We found a significant IR absorption from He atoms
(He-He-He CIA opacity) and a strong pressure distortion of the H$_2$-He 
collision-induced absorption (CIA). We discuss the implication of these results 
for interpretation of the spectra of cool stars.

\end{abstract}

\section{Introduction}
Atmospheres of old and cool white dwarfs deliver valuable information on stellar and planetary evolution.
This is because of relatively simple cooling physics that allows for accurate cosmochronometry \citep{R06,FBB01} and of the stratified 
nature of a white dwarf star, which allows for tracing the content of acreted planetary-like material \citep{F10}.
However, in order to properly asses this information, atmospheres of these stars must be well characterized and understood.
This is often difficult with the standard models that are based mainly of the ideal-gas description of physics and chemistry.
The atmospheres of helium-rich stars reach densities of fluid (up to a few g/cc \citep{K06a,K10,B97}) and there are severe problems in
proper characterization of such old stars (e.g. \citet{GCT15}). Impact of the dense atmospheres on stellar spectra are illustrated by the
pressure distortion of $\rm C_2$ bands \citep{SBF95,B97,K10} and also with some problems with fitting the metal lines profiles \citep{DBL07}.
The former problems in fitting the UV spectra and U and B photometric 
bands of less extreme hydrogen-rich white dwarfs \citep{B97}, solved by introduction of the Ly-$\alpha$ red wing opacities \citep{KS06,K06b},
also show the importance of proper modeling, accurate fitting and characterization of spectra of cool white dwarfs,
for correct understanding of these stars \citep{SHK14}.

In the last two decades various high-pressure improvements have been introduced to the modeling. These include
the non-ideal equation of state \citep{SJ99}, the opacities \citep{IRS02,K10,K14,KS06,K06b}, the refraction \citep{KS04} and the chemical abundances 
of species \citep{K06a}. The further tests of Ly-alpha red wing profiles validated this absorption mechanism for a sample
of cool stars \citep{SHK14}. While the less extreme atmospheres of hydrogen-rich white dwarfs are now well reproduced 
by the current models, the modeling of more extreme helium-rich atmospheres still encounters difficulties \citep{K12,GCT15}. 
In spite of significant work devoted to improvement of high density chemistry and physics description,
no conclusive models describing dense, helium-rich atmospheres exist. In particular, a group of so-called ultracool white dwarfs,
which are believed to be helium-rich stars, have rather vague atmospheric parameters assignment \citep{GCT15}. We suspect that 
the pressure-distortion of IR opacities, namely the collision induced absorption (CIA, \citet{F93}), contributes to much of these difficulties.

Having steady improvement in the performance and availability of the computational power and quantum chemistry simulation software
and techniques, many dense helium problems could be tackled by atomistic simulations \citep{JK14}.
The IR opacities could be directly simulated using {\it ab initio} methods of quantum chemistry \citep{G91,SBP97,JK14}. Recently we attempted 
such simulations and the first results are briefly discussed here in the context of interpreting the near- and mid-IR 
spectra of ultracool stars.

\section{He-He-He CIA opacity}


\begin{figure*}[t!]
\begin{center}
\resizebox{0.6\hsize}{!}{\rotatebox{0}{\includegraphics{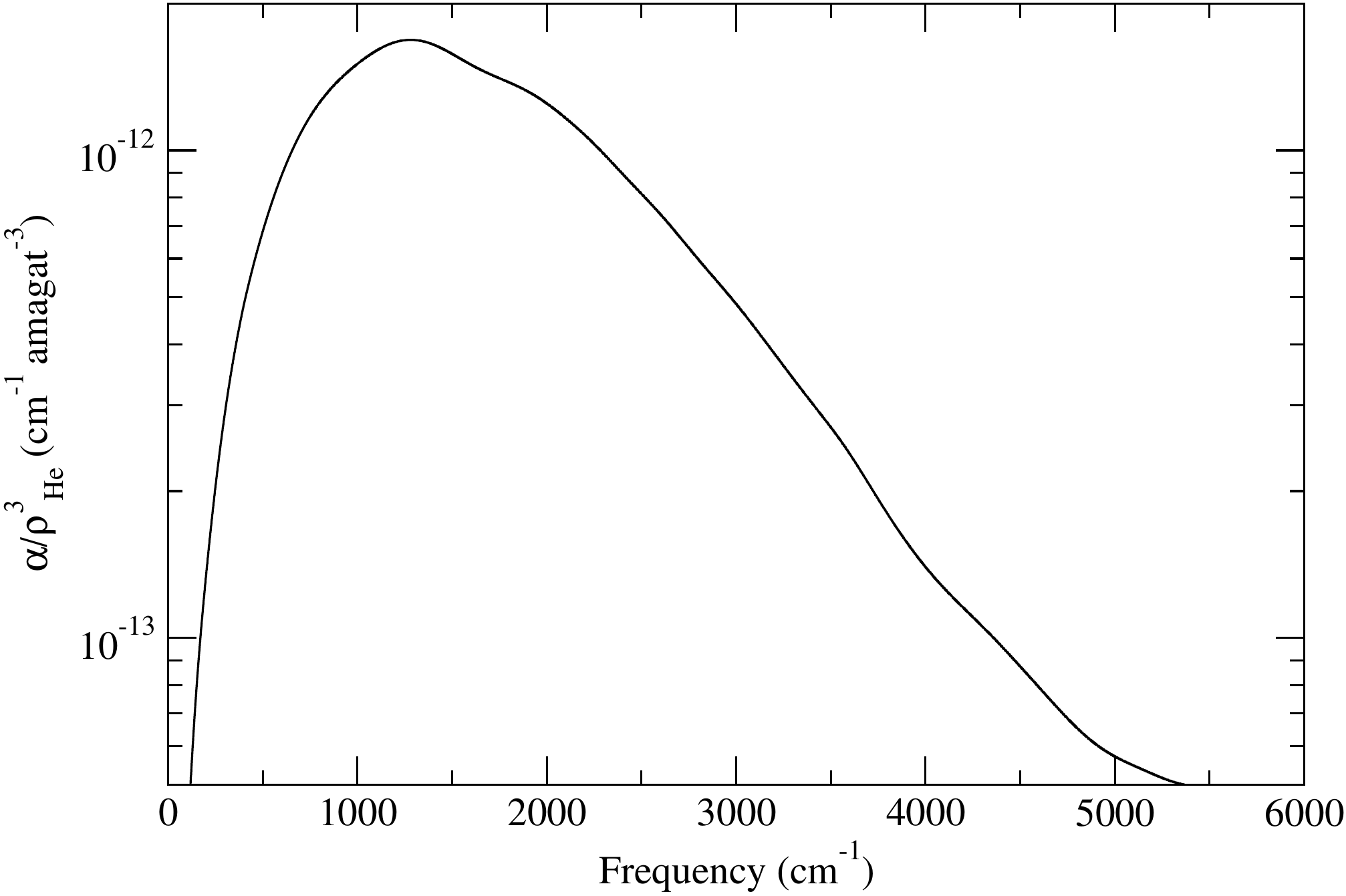}}}
\end{center}
\caption{
Simulated IR absorption profile of dense helium obtained for $T = 5000\rm \, K$ \citep{K14}.
The densities are expressed in amagat = $2.68678 \cdot 10^{19}\, \rm cm^{-3}$.\label{F1}}
\end{figure*}

\begin{figure*}[t!]
\begin{center}
\resizebox{0.6\hsize}{!}{\rotatebox{0}{\includegraphics{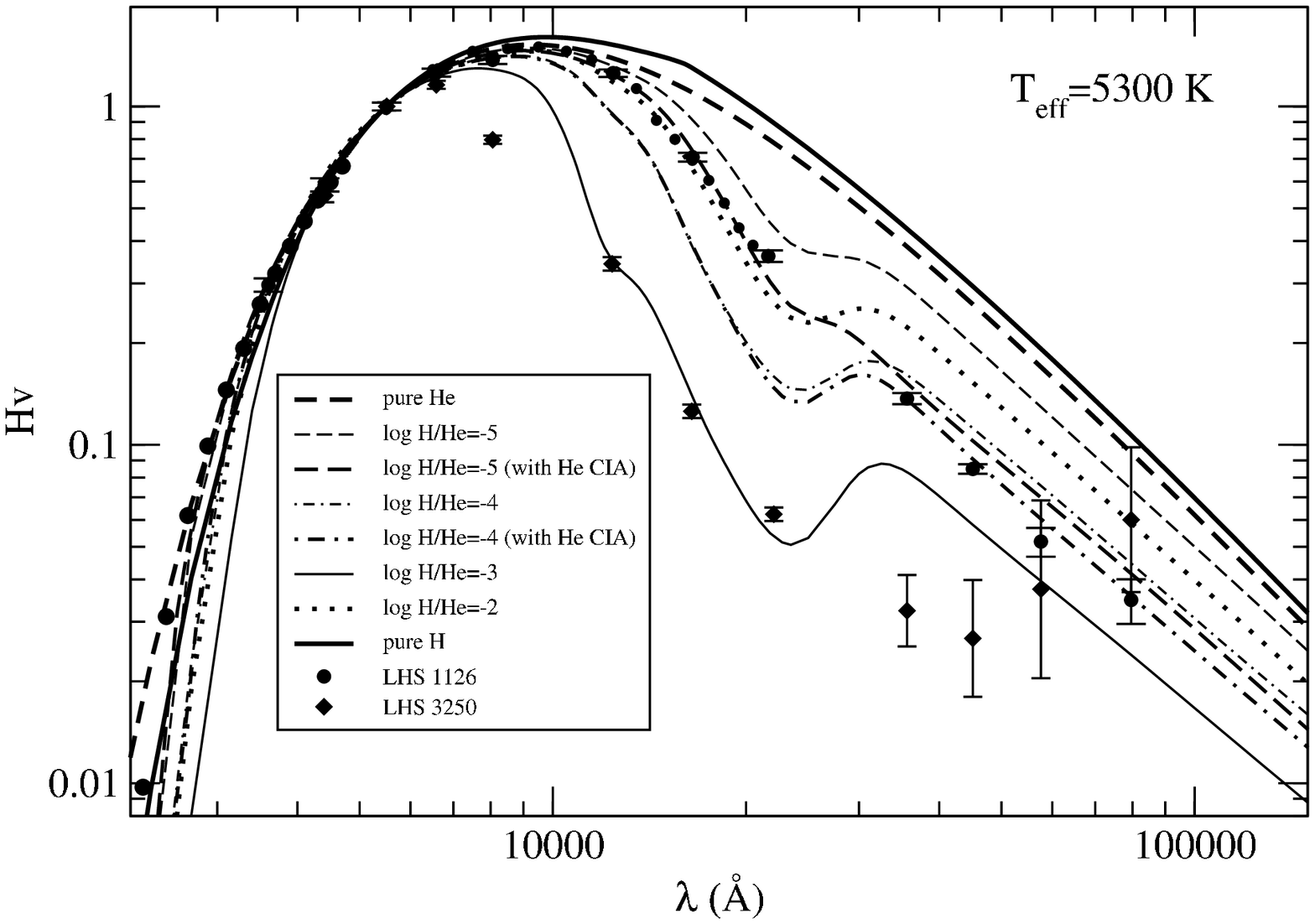}}}
\end{center}

\caption{
Comparison of the SED of the He-rich stars LHS 1126 and LHS 3250 with models of different $\rm H/He$ atmosphere composition, 
$T_{eff}=5300\,$K and $\log g=8$.  The photometric fluxes are from \citet{BLR01}, \citet{KK08} and \citet{KK09}.
For LHS 1126 the UV flux is from  \citet{WKL02} and the near-IR flux is our new data \citep{K12}.
All the spectra are normalized to the $V$ band flux. The thick long dashed and dot-dashed lines represent the models with the He-He-He CIA opacity.\label{F2}
}
\end{figure*}

Recently, using {\it ab initio} molecular dynamics technique, \citet{K14} simulated the CIA opacities of pure, dense helium and found significant absorption
resulting from triple collisions between helium atoms. This absorption 
mechanism was postulated (e.g. \citet{F93}), but had been never before observed experimentally or computed. The obtained absorption coefficient 
for a representative temperature of 5000 K is given in Figure \ref{F1}. Although the intensity of this absorption mechanism is about four orders of magnitude 
lower that the relevant H$_2$/H-He CIA opacities,
because He atoms are the dominant species it becomes a significant IR absorption mechanism in the atmospheres with $\rm He/H>10^4$ (see Fig. \ref{F2}).
When we applied the new absorption mechanism to the spectrum of LHS1128 star, which is broadly discussed by \citet{K12} (Fig. \ref{F2}), we obtained much better 
fit to the IRAC fluxes. This opacity source thus explains a flat-like behavior of the observed IR fluxes, 
as opposite to the $2.3\,\rm \mu m$ flux suppression and the $3.5\,\rm\mu m$ flux excess predicted by the current models (Fig. \ref{F2}).
 

\section{
H$_2$-He CIA opacities}

\begin{figure*}[t!]
\begin{center}
\resizebox{0.6\hsize}{!}{\rotatebox{0}{\includegraphics{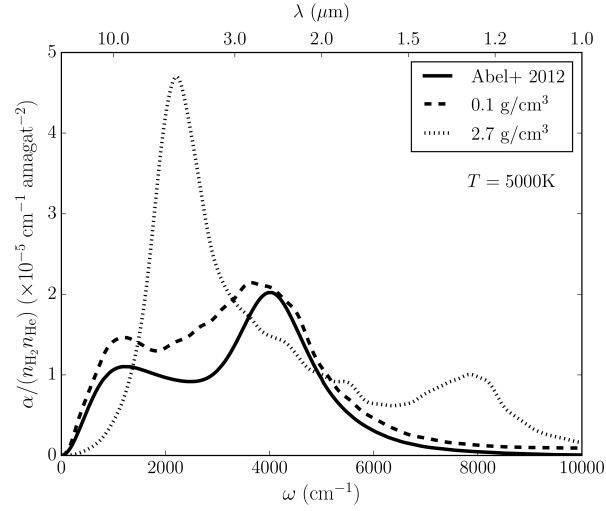}}}
\end{center}
\caption{The comparison of the simulated $\rm H_2-He$ CIA opacities at low and high densities. Note that the results of
\citet{AFL12}, plotted here as a reference, are not corrected for high density (triple collisions, \citet{LS92}),
and should be thus underestimated by $\sim30\%$ at $\rho=0.1\,\rm g/cc$.\label{F3}
}
\end{figure*}

\begin{figure*}[t!]
\begin{center}
\resizebox{0.6\hsize}{!}{\rotatebox{0}{\includegraphics{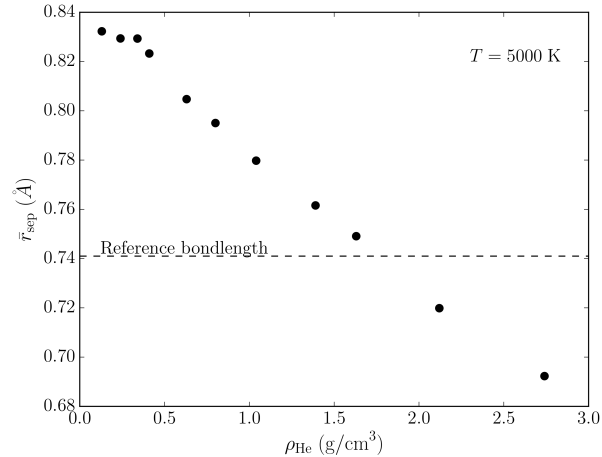}}}
\end{center}
\caption{
The simulated average interatomic separation in H$_2$ molecule as a function of density.\label{F4}
}
\end{figure*}

\begin{figure*}[t!]
\resizebox{0.6\hsize}{!}{\rotatebox{0}{\includegraphics{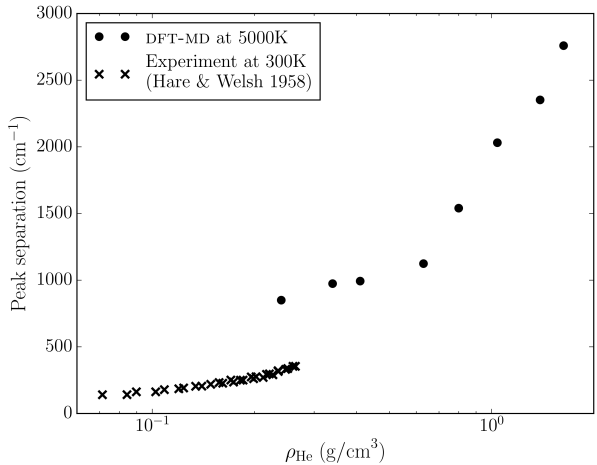}}}
\caption{The main vibrational peak separation as a function of density. The different symbols represent the values simulated here and the experimental results of \citet{HW58}.\label{F5}
}
\end{figure*}

Using the same simulation method as \citet{K14} used for simulation of the He-He-He absorption (see also \citet{BDK16} in this volume) 
we have simulated the $\rm H_2-He$ CIA opacities. As indicated in Fig. 2 of \citet{BDK16} we found a significant distortion 
of this absorption mechanism for densities $\rho>0.1\,\rm g/cc$. The obtained profiles for the low and high densities 
are compared in Figure \ref{F3}. As indicated in the figure, there are three main pressure effects. (1) The rototranslational band peak
at $\sim 1000 \, \rm cm^{-1}$ becomes more pronounced and (2) shifted towards higher frequencies. We interpret this as a result of
triple collisions \citep{LS92} and decrease in the separation between hydrogen atoms in H$_2$ with increase in density, and related decrease of the H$_2$ rotational inertia, 
as illustrated in Fig. \ref{F4}. At the same time, (3) the $\sim 4200 \, \rm cm^{-1}$ main vibrational
band shifts to the higher frequencies and splits into two peaks due to an interference 
between the dipole moments induced in the successive collisions of H$_2$ and He atoms \citep{K68}. This splitting has been experimentally observed before \citep{HW58}
and in Figure \ref{F5} we compare those results with these simulated by us. We obtained larger splitting because 
simulations were performed at much higher temperature, thus with higher thermal collisions energy than in the experiments 
performed at ambient conditions.


The obtained results could be used for the interpretation of difference between the modeled and the observed spectra of 
ultracool white dwarfs, such as LHS3025. Compared to the synthetic spectrum computed using the standard CIA opacity description
of \citet{AFL12}, the observed spectrum does not show the clear minimum at $2.3\rm \, \mu m$ and the flux excess at $\sim 3.5\rm \, \mu m$ 
that are predicted by models, and indicates more absorption in the near-IR. This is consistent with the simulated broadening and the redistribution 
of the main vibrational peak and the enhancement of rototranslational peak. The absorption spectrum simulated at $\rho=2.7\,\rm g/cc$ (Fig. \ref{F3}) indicates 
maximum of absorption at $\rm \sim5\,\rm \mu m$ and more absorption in the near-IR, which is consistent with the observed 
spectral energy distribution of LHS3025, and also other ultracool white dwarfs \citep{GCT15}. More elaborate analysis requires more complete set of simulated spectra,
including investigation of the temperature dependence, derivation of the relevant distortion model, calculations of a grid of atmosphere models 
and their usage for subsequent fitting of spectra and analysis of ultracool white dwarfs. This work is currently in progress.

\section{Conclusion}
We have showed the results of simulations of He-He-He and H$_2$-He CIA infrared opacities and their 
impact for the interpretation of spectra of cool, helium-rich stars, including ultracool white dwarfs.
We discuss the impact of the newly found He-He-He CIA opacity from pure helium on the spectra.
By improved fit to the spectrum of LHS1128 star we show that this opacity mechanism could be responsible for 
a flat-like infrared spectral energy distribution of these stars, which could not be reproduced with previous models.
The subsequent simulations of H$_2$-He CIA opacities show significant pressure-distortion of this absorption mechanism
with the simulated intensity change being consistent with the one deducted from the mismatch between the current models prediction
and the observed spectral energy distribution for star LHS3250. However, more complete simulations of the IR opacities and subsequent computation of new models 
are required for complete analysis and refitting the spectra of cool, helium-rich stars, which is currently in progress.
Last but not least, the  simulations conducted here and the subsequent analysis show that the modern atomistic modeling methods could be successfully applied 
to modeling condensed stellar atmospheres and that, on the other hand, the atmospheres of white dwarfs
represent excellent laboratories for testing the physics and chemistry of matter at extreme conditions. 

\acknowledgements S.B. acknowledges support from NSERC (Canada), FRQNT (Qu\'ebec) and DAAD (Germany).

\end{document}